\documentclass[twocolumn,superscriptaddress,floatfix,prl,showpacs]{revtex4-1}
\usepackage{graphicx}
\usepackage{amsmath}
\usepackage{amssymb}
\usepackage{color}
\usepackage{braket}
\usepackage{dsfont}
\usepackage{natbib}
\usepackage{mathtools}

\begin{document}
\title{Topological phase transition in quantum spin Hall insulator in the presence of charge lattice coupling}

\author{L.~M.~Cangemi}
\affiliation{SPIN-CNR and Dip. di Scienze Fisiche - Universit\`a di Napoli Federico II - I-80126 Napoli, Italy}

\author{A.~S.~Mishchenko}
\affiliation{RIKEN Center for Emergent Matter Science (CEMS),  
2-1 Hirosawa, Wako, Saitama, 351-0198, Japan}
\affiliation{NRC ``Kurchatov Institute", 123182, Moscow, Russia}

\author{N.~Nagaosa}
\affiliation{RIKEN Center for Emergent Matter Science (CEMS),
2-1 Hirosawa, Wako, Saitama, 351-0198, Japan}
\affiliation{Department of Applied Physics and Quantum-Phase Electronics Center, University of Tokyo, Bunkyo, Tokyo 113-8656, Japan} 

\author{V. Cataudella}
\affiliation{SPIN-CNR and Dip. di Scienze Fisiche - Universit\`a di Napoli Federico II - I-80126 Napoli, Italy}

\author{G. De Filippis}
\affiliation{SPIN-CNR and Dip. di Scienze Fisiche - Universit\`a di Napoli Federico II - I-80126 Napoli, Italy}

\begin{abstract}
  By using the cluster perturbation theory, we investigate the effects of the local electron-phonon interaction in the quantum spin Hall topological insulator described by the half-filled Kane-Mele model on an honeycomb lattice. Starting from the topological non trivial phase, where the minimal gap is located at the two inequivalent Dirac points of the Graphene, $\text{K}$ and $\text{K}'$, we show that the coupling with quantum phonons induces a topological-trivial quantum phase transition through a gap closing and reopening in the $\text{M}$ point of the Brillouin zone. The average number of fermions in this point turns out to be a direct indicator of the quantum transition pointing out a strong hybridization between the two bare quasiparticle bands of the Kane-Mele model. By increasing the strength of charge-lattice coupling, the phonon Green's propagator displays a two peak structure: the one located at the lowest energy exhibits a softening that is maximum around the topological transition. Numerical simulations provide also evidence of several kinks in the quasiparticle dispersion caused by the coupling of the electrons with the bosonic lattice mode.     
  
\end{abstract}
\maketitle

Topological insulators are one of the most active field of research in modern condensed matter physics. These materials are band insulators exhibiting a non-trivial topological band structure \cite{Hasan,QiZhang,ber1}. The distinct topological phases are described in terms of the values of a bulk invariant \cite{ber1} determining the presence of gapless chiral edge modes, along with insulating behavior in the bulk. 

The Integer Quantum Hall Effect (IQHE) was first described in terms of topological Chern number \cite{Th}: the value of this invariant describes the perfectly quantized transverse Hall conductance observed in the experiments \cite{Von}. Following the seminal paper by Haldane \cite{hal}, who introduced a tight binding model on the honeycomb lattice with broken Time Reversal (TR), it was realized that IQHE can occur even in the absence of a net magnetic flux, provided that TR symmetry is broken: this prototypical system belongs to the class of Chern insulators \cite{Altland,sch,kit,Kruthoff}. 

The field experienced an enormous progress following the works by Kane and Mele \cite{Kane1,Kane2}, Bernevig and Zhang \cite{ber}, which first theoretically predicted the existence of TR symmetry-protected topological band insulators in 2D, known as Quantum Spin Hall (QSH) insulators. In these systems, the possible topological phases are classified according to values of a $\mathbb{Z}_2$ invariant \cite{Kane2,moore,FuKane}; TR symmetry-protected helical (opposed chirality) pairs of edge modes are present and their number is related to the change in the value of $\mathbb{Z}_2$ invariant at the edge by bulk-boundary correspondence \cite{Kane2,Hasan,ber}. QSH insulators were found to belong to $\textmd{DII}$ topological class \cite{Altland,sch,kit,Kruthoff}. Shortly after these works, the theory of QSH insulators was generalized to 3D \cite{Fu3D}, and it was confirmed by a number of experimental findings \cite{Konig,Hsieh3D}.

While topological insulators are described by one-particle band theories, in recent years a great deal of work has been done to study the influence of electronic correlations in these systems. Various sophisticated formulations of topological invariants for the interacting case have been developed \cite{Gurarie1,Gurarie2,Wang}: some of them allow the computation of the invariants starting from the Green's functions of the interacting system. During the last decade, a number of works on correlated topological insulators have focused on the effects of electron-electron (e-e) interactions: choosing well-estabilished bare topological band models as starting points, e-e interaction, modeled by means of Hubbard-like terms, have been considered \cite{rac,Hohen}. A thorough study of the stability of the topological insulating phase, as well as the full characterization of the phase diagrams in the presence of correlations, have been performed\cite{RachLeHur,Zheng,Hohen2}. Furthermore, the search for interaction-induced topological phases, both with and without band-structure analogues \cite{Raghu}, has been carried out, and the existence of much more exotic topological phases has been proposed, e.g. fractional topological insulators\cite{Neupert,Mac}.     

Although the physics of topological insulators in the presence of Coulomb interactions is settled up to the level of finding better approximations or establishing exact phase borders by approximation-free methods \cite{prok}, the unavoidable electron-phonon coupling (EPC) in real materials requires investigation of its effects on the topological properties of all prototypical non-interacting models. Besides, in contrast to instantaneous Coulomb coupling, the EPC induces a retarded interaction with propagator having its own (phonon) frequency. This feature can be used to probe the typical energy scales involved into the formation of topological phase. Influence of EPC was studied both in the context of Graphene and Chern insulators \cite{Hohen3,Can}, and in strongly correlated bosonic systems on dynamical lattices \cite{Gonz}. Furthermore, recent works \cite{mona} have shown that a different model for EPC, dependent both on quasi-particle and phonon momentum, can stabilize type II Dirac-Weyl cones in Lieb lattices, i.e. non trivial topological band structure due to polaron bands can occur.

Below, we study the effect of EPC on the topological properties of the Kane-Mele (KM) model of a QSH band insulator. It describes spin $1/2$ fermions in the presence of a hopping contribution, an intrinsic spin-orbit coupling (SOC) term, and an additional Rashba SOC on a half-filled honeycomb lattice \cite{Kane1,Kane2}. In this paper we consider the Rashba spin-orbit interaction to be zero. The Hamiltonian can be obtained by considering two copies of the model introduced earlier by Haldane \cite{hal}, where the second-nearest neighbor hoppings for spin-$\uparrow$ and -$\downarrow$ electrons are complex valued and complex conjugate to each other. We will focus our attention on the case where next-nearest-neighbor spin-orbit hopping integral is purely imaginary, i.e. in the presence of the particle-hole symmetry. Furthermore we describe the lattice dynamics by introducing on-site optical modes, coupled to fermions by means Holstein-type interaction term \cite{Can}.        

We calculate the single particle Green's propagator in the thermodynamic limit by using the cluster perturbation theory (CPT). We choose the model parameter values such that the minimal gap of the bare topological insulator is located at the two inequivalent Dirac points of the Graphene, $\text{K}$ and $\text{K}'$, and we show that EPC induces a topological-trivial quantum phase transition through a gap closing and reopening in the $\text{M}$ point of the Brillouin zone. By following Wang et al. \cite{Wang} we compute the topological invariant via the parity eigenvalues of the fully interacting Green's function obtained at the time-reversal invariant momenta and zero energy. The numerical simulations show that, by varying the strength of EPC, the $\mathbb{Z}_2$ invariant changes from one to zero just where the gap closes. Here a strong hybridization between the two bare quasiparticle bands of the KM model occurs. We show also that the average number of fermions at the $\text{M}$ point of the Brillouin zone can be used as a specific indicator of the quantum phase transition. Furthermore: i) a splitting and a softening of the phonon Green's function are observed around the topological transition revealing the energy scale stabilizing the topological phase; 2) many kinks in the electron renormalized dispersion appear as direct consequence of the coupling between the charges and the lattice boson mode.  

{\it The model.} The KM model \cite{Kane1,Kane2} describes spin-$1/2$ fermions on a honeycomb lattice, it is a prototypical model of QSH. It can be described in terms of two copies of the Haldane model, one for each value of the spin along $\hat{z}$ axis, with additional Rashba SOC term. Below, we take into account only the SOC component along $\hat{z}$, thus disregarding Rashba-type terms (they break the spin symmetry and the spatial inversion symmetry), and we fix the mass term of the Haldane model to be zero. Then the KM Hamiltonian reads:     
\begin{equation}\label{eq:KaneMele}
  H_{KM}=-t_1\sum_{\mathclap{<i,j>,\sigma}}c_{i\sigma}^{\dagger}c_{j\sigma}-\sum_{\mathclap{\ll i,j\gg,\sigma}}t_{so}e^{i\xi_{i\sigma}\phi}c_{i\sigma}^{\dagger}c_{j\sigma}
\end{equation}
where $c_{i\sigma}^{\dagger}(c_{i\sigma})$ are fermionic creation (annihilation) operators, on the site $i$ with spin $\sigma$, on the two sublattices $\left (A,B \right)$, $t_{1}$ is the nearest-neighbor hopping element, $t_{so}$ is the strength of the second nearest neighbor SOC term, $\xi_{i\sigma}=\pm 1(\mp 1)$ respectively for hopping on the sublattice sites $\left (A,B \right)$ and spin-$\uparrow$ (spin-$\downarrow$). As evident from \eqref{eq:KaneMele}, the phase choice in the SOC term ensures that the net magnetic flux through the honeycomb is zero. However, due to the fact that the phases for different spin values are complex-conjugate of each other, the SOC term can open a gap in the high-symmetry points of the first Brillouin zone (BZ) of the honeycomb lattice without breaking TR symmetry. In what follows, we restrict to the case of particle-hole symmetry, i.e. we take $\phi=\pi/2$ and fix the chemical potential to be zero.        
The KM model can be rewritten in the quasi-momentum space, introducing creation (annihilation) fermionic operator $(a_{k\sigma},b_{k\sigma})$ on each sublattice 
\begin{equation}
  H_{KM}=\sum_k \Psi^{\dagger}_k\mathcal{H}(\boldsymbol{k})\Psi_k
  \end{equation}
with $\Psi^{\dagger}_k= \begin{pmatrix}a^{\dagger}_{k\uparrow} & b^{\dagger}_{k\uparrow}& a^{\dagger}_{k\downarrow}& b^{\dagger}_{k\downarrow} \end{pmatrix}$ and $\mathcal{H}(\boldsymbol{k})$ denotes the four-dimensional matrix $\mathcal{H}(\boldsymbol{k})=  \boldsymbol{h}(\boldsymbol{k})\cdot{\boldsymbol{\sigma}}\boldsymbol{\mathds{1}}_{\sigma \sigma^{\prime}}+h_{so}(\boldsymbol{k})\sigma_{z}(s_{z})_{\sigma \sigma^{\prime}}$, where $s_z$ is the spin operator along $\hat{z}$. As usual, we denote with  $\boldsymbol{\sigma}=\sigma_i\mbox{, } i=x,y,z$ Pauli matrices, and the quasi-momentum $\boldsymbol{k}=(k_x\mbox{, }k_y)$ belongs to BZ. The vector $\boldsymbol{h}(\boldsymbol{k})$ is generally analogous to a 3D magnetic field, $\boldsymbol{h}(\boldsymbol{k})=(h_x(\boldsymbol{k}),h_y(\boldsymbol{k}),h_z(\boldsymbol{k}))$; in the presence of inversion symmetry, the field component along $\hat{z}$ vanishes. The matrices $(\mathds{1},s_{z})$ act on the space of the spin degree of freedom. The 4D Hamiltonian $\mathcal{H}(\boldsymbol{k})$, which is invariant under TR and inversion symmetry, can be written as a linear combination of $\text{SU}(4)$ matrices $\Gamma_{\alpha}=(\sigma_{x}\mathds{1},\sigma_{y}\mathds{1},\sigma_{z}s_{x},\sigma_{z}s_{y},\sigma_{z}s_{z})$ \cite{ber1,FuKane,RachLeHur}. It can be readily diagonalized, through a unitary transformation, to give: $\mathcal{H}(\boldsymbol{k})=E_{k,-}\sum_{\sigma}\gamma^{\dagger}_{k,-,\sigma} \gamma_{k,-,\sigma}+E_{k,+}\sum_{\sigma}\gamma^{\dagger}_{k,+,\sigma} \gamma_{k,+,\sigma}$, where $\gamma_{k,\pm,\sigma}$ denote quasi-particle creation (annihilation) operators and $E_{k,\pm}=\sqrt{h^2_{so}(\boldsymbol{k}) +h^2_x(\boldsymbol{k})+ h^2_y(\boldsymbol{k})}$ are the two-fold degenerate quasi-particles energy bands. The SOC term, which is $\propto \Gamma_{5}$, opens a gap at Dirac points $\text{K},\text{K}^{\prime}$. As long as the gap is open, the topological phases of KM can be classified according to the values of $\mathbb{Z}_2$ invariant. We aim at describing the effects of the EPC on the topological properties of KM model \eqref{eq:KaneMele}. Following a recent work \cite{Can}, we introduce the EPC by means of the Holstein model, which linearly couples the charge fluctuations to the displacement of on-site lattice vibrations:
\begin{equation}\label{eq:Interacting}
  H=H_{KM} + \omega_{0}\sum_i d^{\dagger}_i d_i + g\omega_{0}\sum_i(n_i-1)(d^{\dagger}_i + d_i)
\end{equation}

\begin{figure}[thb]
  \includegraphics[width=0.99\columnwidth]{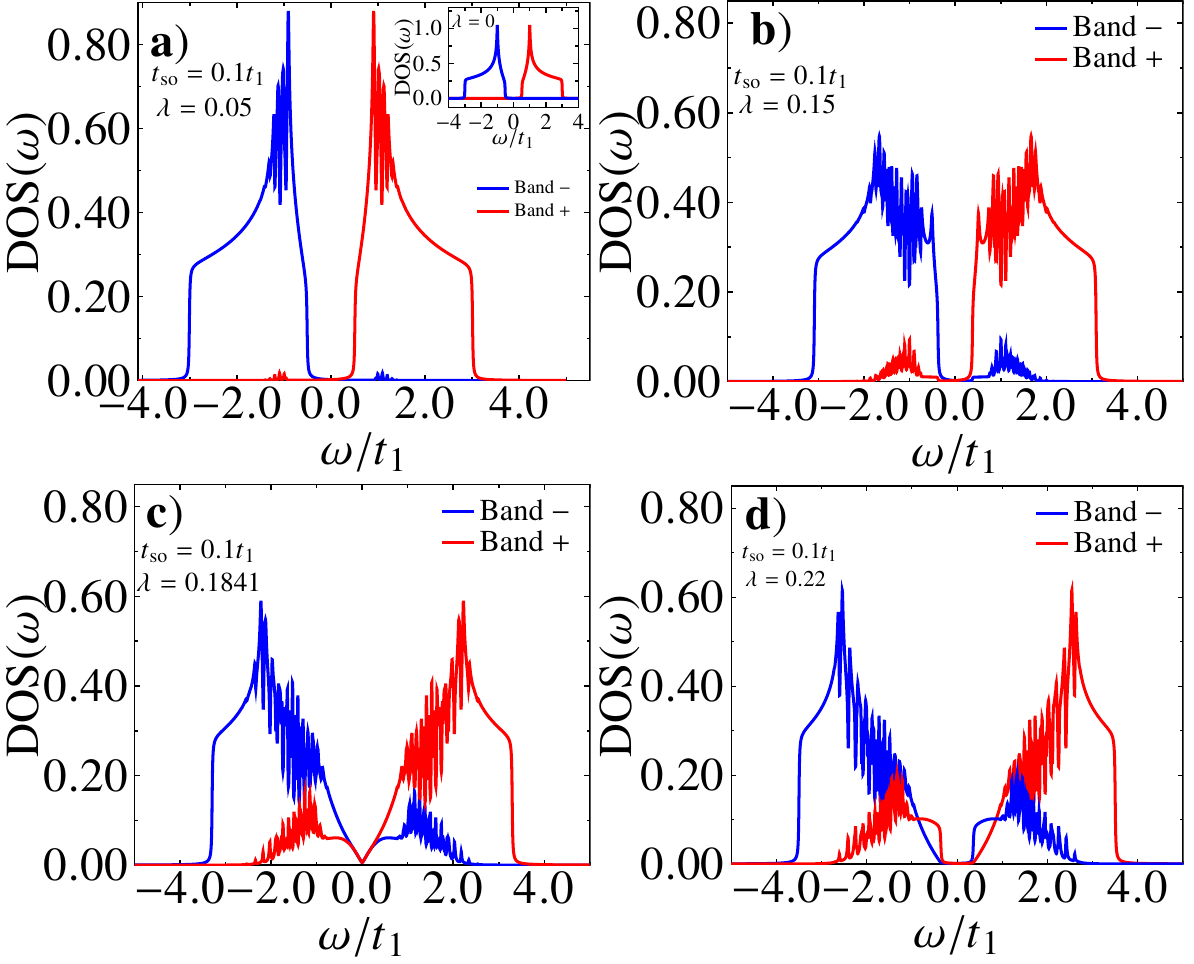}
  \caption{\label{fig:1} (color online)
    Density of states for different values of EPC.}
\end{figure}

We employ shorthand notation $d^{\dagger}_i$($d_i$) for two different bosonic operators, which respectively create (annihilate) a phonon on the two $\left (A,B \right)$ sublattice sites, $n_i$ indicates the electron number operator on the site $i$, $\omega_{0}$ is the optical mode frequency, and $g$ represents the strength of the coupling with lattice. We introduce also the dimensionless parameter $\lambda=g^2 \omega_0/4 t_1$. Here we restrict our attention to the case of half-filling, i.e. $\sum_{i \in A} a^{\dagger}_i a_i + \sum_{i \in B} b^{\dagger}_i b_i =2N_c=N$, where $N_c$ ($N$) is the number of unit cells (lattice sites). We choose $t_1=1$, $\omega_0=0.1$, $t_{so}=0.1$, and use units such that $\hbar=1$.

{\it The results.} The single particle Green's function $\boldsymbol{G}_{\sigma}(\boldsymbol{k},z)$, where $z=\omega+i\eta$ lies in the complex upper half plane, is obtained by using the CPT. Due to the explicit $s_z$ conservation of the Hamiltonian, the Green's function is block-diagonal in spin-space, and, furthermore, $\boldsymbol{G}_{\sigma}$ is a $2$x$2$ matrix in the $\left (A,B \right)$ sublattice basis. We emphasize that this does not mean that up and down spin electrons are decoupled. Indeed they interact with each other through the EPC. By using the unitary transformation introduced to diagonalize the Hamiltonian, it is straightforward to extract the Green's functions relative to the operators $\gamma_{\boldsymbol{k},\pm,\sigma}$ representing the quasiparticles in the absence of the EPC. In Fig.~\ref{fig:1} we plot, for different values of the charge-lattice coupling, the density of states: $DOS_{(\mp,\mp)}(\omega)=\frac{1}{N_c}\sum_{\boldsymbol{k}} A_{\sigma,(\mp,\mp)}(\boldsymbol{k},\omega)$, where the two spectral weight functions $A_{\sigma,(\mp,\mp)}(\boldsymbol{k},\omega)$ are given by: $A_{\sigma,(\mp,\mp)}(\boldsymbol{k},\omega)=-\frac{\Im{G_{\sigma,(\mp,\mp)}(\boldsymbol{k},z)}}{\pi}$. At $\lambda=0$, where $A_{\sigma,(\mp,\mp)}$ are delta functions peaked at $E_{k,-}$ and $E_{k,+}$, the two density of states exhibit Van Hove singularities and a finite gap due to the presence of spin-orbit coupling $t_{so}$. In this case the two bands are completely separated, i.e. $A_{\sigma,(-,-)}$ ($A_{\sigma,(+,+)}$) is different from zero only at $\omega$ lower (greater) than $\mu$. By increasing EPC the two bare bands of the KM model hybridize and the gap reduces.

\begin{figure}[thb]
  \includegraphics[width=0.99\columnwidth]{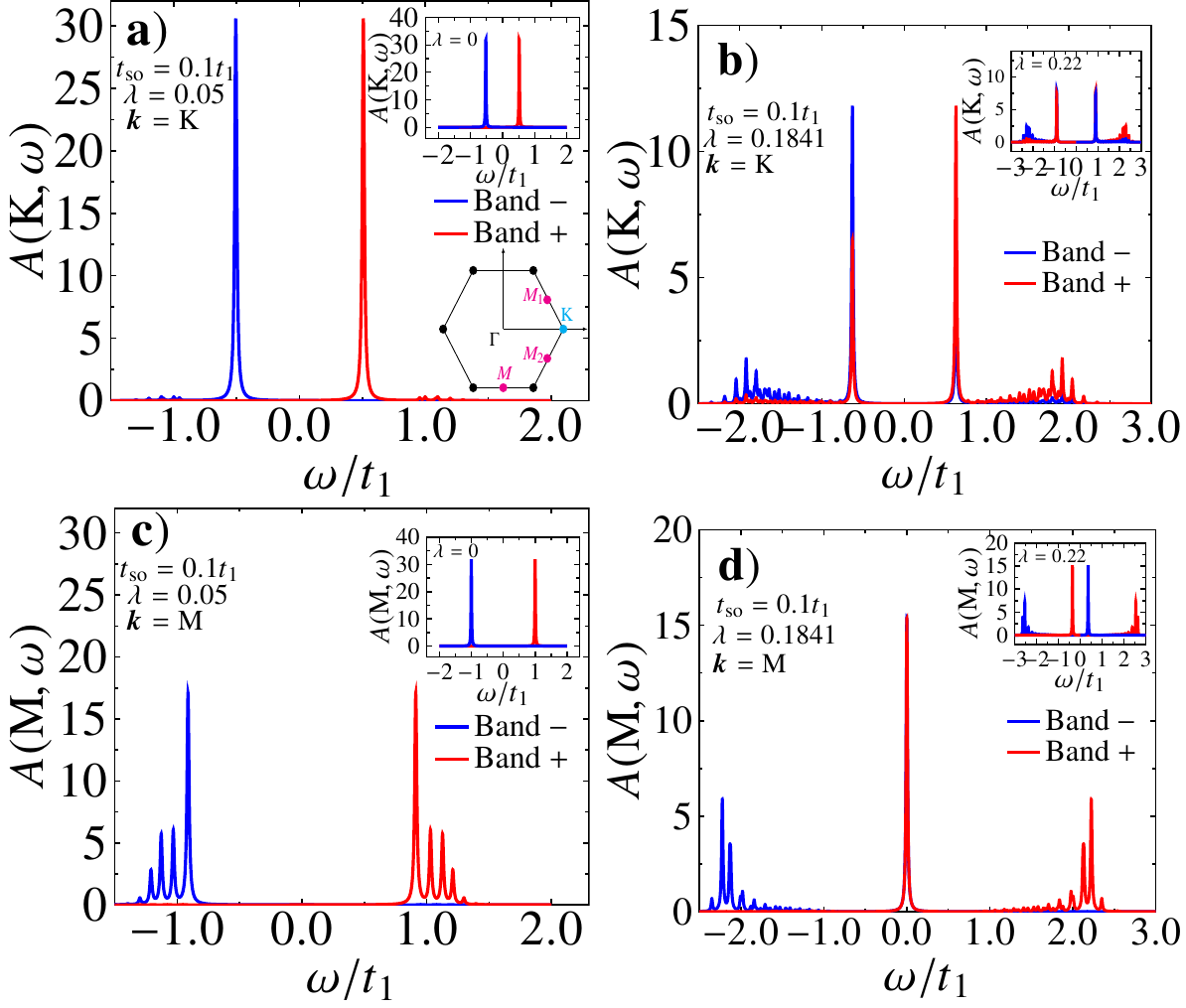}
  \caption{\label{fig:2} (color online)
    Spectral weight function at: $\text{K}$ ((a) and (b)) and $\text{M}$ ((c) and (d)) for different values of $\lambda$ across the quantum phase transition. In the panel (a) the Brillouin zone with the time reversal invariant momenta ($\Gamma$, $\text{M}$, $\text{M}_1$,$\text{M}_2$) and $\text{K}$ point indicated.}
\end{figure}

\begin{figure}[thb]
  \includegraphics[width=0.99\columnwidth]{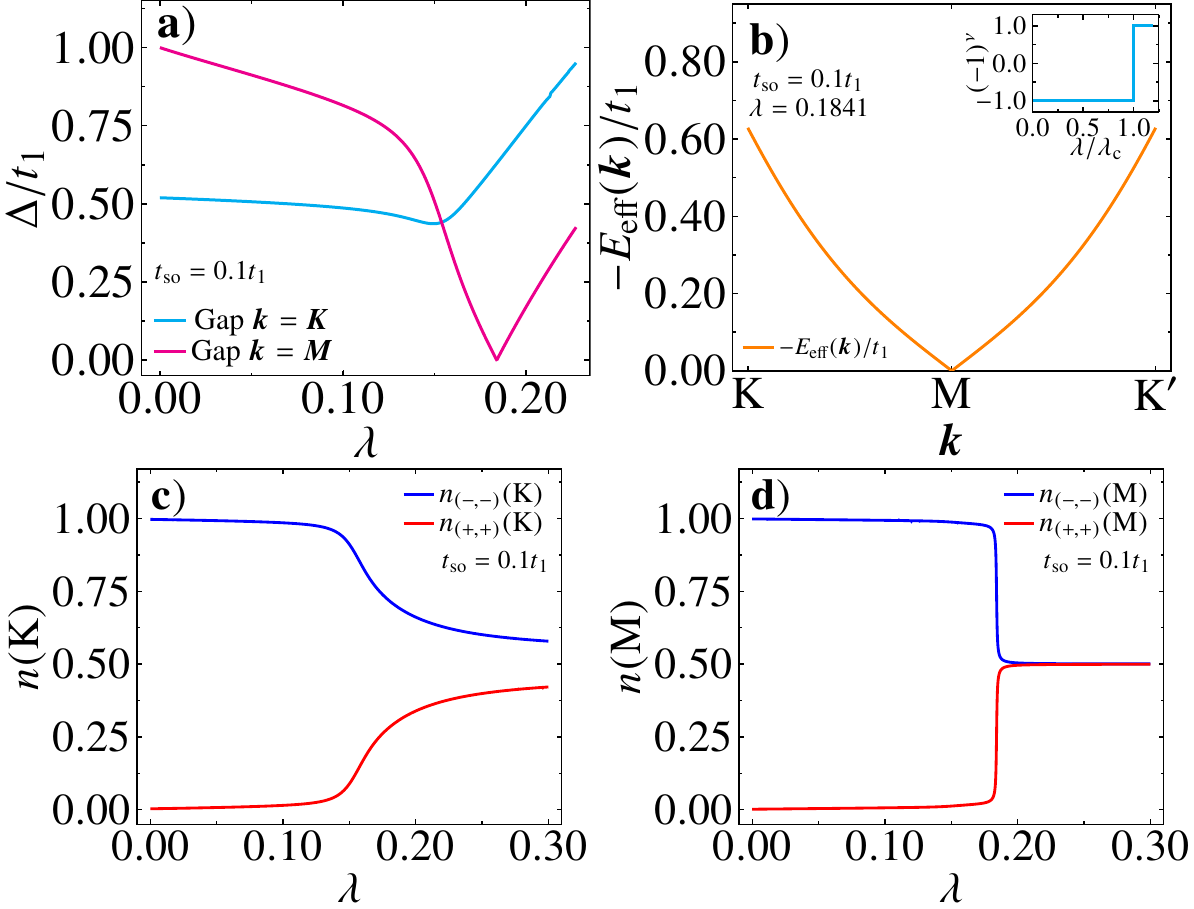}
  \caption{\label{fig:3} (color online)
 (a): the gap at $\text{K}$ and $\text{M}$ as function of $\lambda$; (b): the quasiparticle dispersion, within the hole sector, along the $\text{K}$-$\text{M}$-$\text{K}'$, at $\lambda_c$; $n(\text{K})$ and  $n(\text{M})$ ((c) and (d)) as function of EPC.}
\end{figure}

In particular the spectra point out that, near the Van Hove singularity, there is a stronger effective charge-lattice coupling. At $\lambda_c=0.1841$, the gap closes and, then, by further increasing EPC, reopens and becomes larger and larger. Figures \ref{fig:2}a and \ref{fig:2}b show that, at the $\text{K}$ point, the spectral weight functions exhibit a finite gap for any value of $\lambda$, i.e. the Dirac points of the Graphene, $\text{K}$ and $\text{K}'$, are not responsible of the behavior observed in the $DOS$. On the other hand, the plots in Fig.~\ref{fig:2}c, Fig.~\ref{fig:2}d and Fig.~\ref{fig:3}a point out that, by increasing the strength of EPC, the gap at $\text{M}$ point reduces, closes exactly at $\lambda_c$ (here both the functions display a peak at $\mu$) and reopens for $\lambda > \lambda_c$. In particular, at $\lambda_c$, we followed, within the hole sector and near $\text{M}$ point, the dispersion of the lowest energy quasiparticle peak. Fig.~\ref{fig:3}b shows that, at the $\text{M}$ point, a semimetal Dirac cone appears just at $\lambda_c$. It is also worth mentioning the behavior of the average number of electrons, $n_{(-,-)}$ and $n_{(+,+)}$, associated to the lower and upper bands of the bare KM model at the $\text{M}$ and $\text{K}$ points. They are obtained integrating the corresponding spectral weight function up to the chemical potential. Fig.~\ref{fig:3}c and Fig.~\ref{fig:3}d show that: i) $n_{(-,-)}(\text{M})$ and $n_{(+,+)}(\text{M})$ present a sharp discontinuity at $\lambda_c$ pointing out that they can be used as direct indicators of the quantum phase transition; ii) $n_{(-,-)}(\text{K})$ and $n_{(+,+)}(\text{K})$ exhibit a change of the sign of the second derivative at the value of $\lambda$ where the gap at $\text{M}$ point becomes less than the one at $\text{K}$ point (see also Fig.~\ref{fig:3}a). Furthermore the value, near $0.5$, reached by all the average fermion numbers for $\lambda > \lambda_c$, points out the strong hybridization, induced by EPC, between the two bare bands of the bare KM model.

In order to shed light on the nature of the observed phase transition, we calculated the $\mathbb{Z}_2$ topological index, that, in two dimensions and in the presence of time-reversal invariance, characterizes topological band insulators. The $\mathbb{Z}_2$ invariant, $\nu$, in the presence of inversion symmetry, can be computed via the parity eigenvalues of the interacting Green's function, obtained within the CPT, at zero energy and the time-reversal invariant momenta $\Gamma_i$. Here: i) $\Gamma_i$ satisfies the relation: $-\Gamma_i=\Gamma_i+\boldsymbol{b}$, where $\boldsymbol{b}$ is a reciprocal-lattice translation vector; ii) the parity operator, i.e. the operator that interchanges the two sublattices and squares to the identity, is represented, in the sublattice basis, by the first Pauli-matrix, $P=\sigma_x$. Indeed, simultaneously diagonalizing the two matrices $P$ and $G_\sigma(\Gamma_i,0)$, and considering, for each of the four momenta $\Gamma_i$, the eigenvalue of $P$, $\delta_i$, for the common eigenvector with a positive eigenvalue of $\boldsymbol{G}_\sigma(\Gamma_i,0)$, it can be shown \cite{Wang} that: $(-1)^{\nu}=\prod_{i=1}^{4}\delta_i$. A topological non-trivial phase ($\nu=1$) is associated with the occurrence of a TR-symmetry-protected pair of gapless edge modes on each edge, which takes quantized conductance. The inset in Fig.~\ref{fig:3}b points out that $\nu$ changes from one to zero just where the gap closes, i.e. at $\lambda_c$. It confirms that at $\lambda_c$ a quantum phase transition, from topological to trivial insulator, occurs. 

Now we investigate the effects of the quantum transition on the lattice. To this aim, we emphasize that the exact integration of the phonon degrees of freedom, through path integral technique, leads to a retarded electron-electron interaction on the same sublattice ($V^{0}$). On the other hand, the effective interaction between two charge carriers obeys the Dyson equation \cite{mahan, fetter}:
\begin{eqnarray}
  V^{eff}_{i,j}(\boldsymbol{q},z)=V^{0}_{i,j}(\boldsymbol{q},z)+V^{0}_{i,h}(\boldsymbol{q},z)\Pi^{*}_{h,k}(\boldsymbol{q},z\
  )V^{eff}_{k,j}(\boldsymbol{q},z),
  \nonumber
\end{eqnarray}
which defines the proper polarization insertion $\Pi^{*}_{i,j}(\boldsymbol{q},z)$. Here $i$ represents a pair of indexes: the first one indicates the sublattice and the other one the spin. At the lowest order $\Pi^{*}_{i,j}(\boldsymbol{q},z)$ is the particle-hole bubble. The next step is to replace, in this lowest order diagram, the unperturbed electron Green's functions with the interacting Green's functions calculated within the CPT. This procedeure allows to obtain the effective interaction between two electrons and, then, the renormalized phonon propagator $D_{i,j}$.

\begin{figure}[thb]
  \includegraphics[width=0.99\columnwidth]{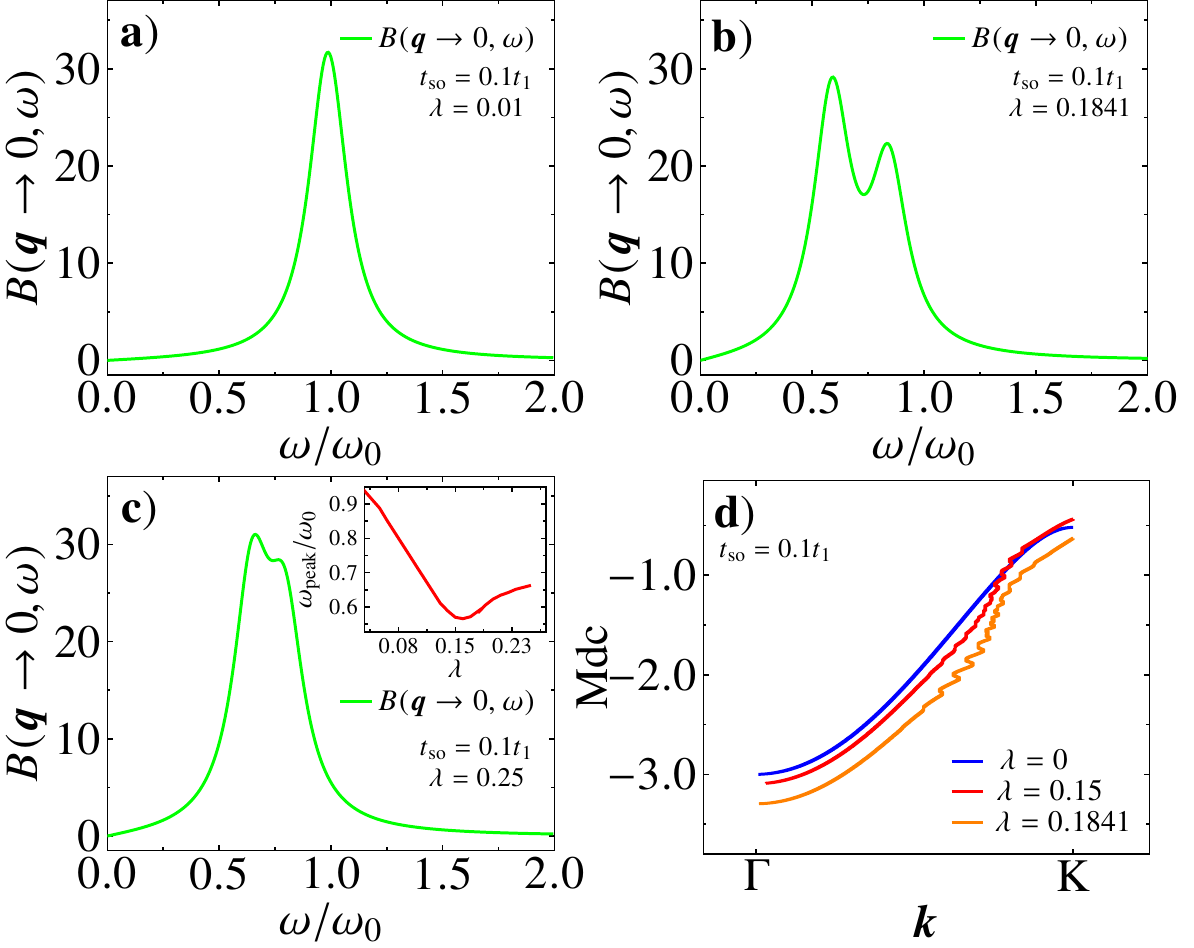}
  \caption{\label{fig:4} (color online)
  Phonon spetral weight function at different values of $\lambda$ ((a), (b), and (c)). In the inset the behavior of the lowest energy peak as function of $\lambda$; (d): hole dispersion from momentum distribution curves along $\Gamma-\text{K}$ direction.}
\end{figure}

We focus our attention on the spectral weight function $B_{(i,i)}(\boldsymbol{q} \rightarrow 0,\omega)=-\frac{\Im{D_{(i,i)}}(\boldsymbol{q},z)}{\pi}$ ($i=(A,\uparrow)$), an odd function, that, in the absence of EPC, is peaked at $\omega=\omega_0$. In Fig.~\ref{fig:4} we show that, by increasing $\lambda$, the spectral weight function displays a two-peak structure. Furthermore: i) the lowest energy peak softens with EPC; ii) the maximum softening occurs across the quantum phase transition (see inset of Fig.~\ref{fig:4}c). It is well known that the doubling of the phonon peak occurs when the phonon frequency is close to electronic excitation which is strongly coupled to phonon \cite{KiMi,Citro}. Hence, the retarded nature of the interaction induced by EPC with its own (phonon) frequency gives us a chance, which is absent in case of instantaneous Coulomb interaction, to pin down the characteristic energy scale stabilizing the topological phase, close to phonon energy within our set of parameters. Finally, in Fig.~\ref{fig:4}d we plot the hole peak dispersion, derived from the momentum distribution curves, along the $\text{K}$-$\Gamma$ direction of the Brillouin zone, at different values of $\lambda$. It is evident the presence of many kinks, at $\lambda \ne 0$, caused by the coupling between the charges and the lattice vibrations.   

{\it Conclusion.} We have highlighted the effects, induced by a local electron-phonon interaction, on the QSH topological insulator described by the half-filled KM model on an honeycomb lattice. By increasing EPC, a quantum phase transition, from topological to trivial insulator, is observed through a gap closing and reopening in the $\text{M}$ point of the Brillouin zone. Here, when the gap is closed, a semimetal Dirac cone appears and a strong hybridization between the two bare quasiparticle bands of the KM model occurs. The abrupt change in the average number of fermions at the $\text{M}$ point, the two peak structure and the softening of the phonon Green's function, and the presence of several kinks in the renormalized quasiparticle dispersion are other distinctive features of this topological quantum phase transition.




\begin{thebibliography}{10}
\providecommand{\url}[1]{{#1}}
\providecommand{\urlprefix}{URL }
\expandafter\ifx\csname urlstyle\endcsname\relax
  \providecommand{\doi}[1]{DOI \discretionary{}{}{}#1}\else
  \providecommand{\doi}{DOI \discretionary{}{}{}\begingroup
  \urlstyle{rm}\Url}\fi  


\bibitem{Hasan}
   M.~Z.~Hasan and C.~L.~Kane, Rev. Mod. Phys. \textbf{82}, 3045 (2010).

\bibitem{QiZhang}
X.~L.~Qi and S.~C.~Zhang, Rev. Mod. Phys. \textbf{83}, 1057 (2011).

\bibitem{ber1}
B.~A.~Bernevig, {\it Topological insulators and superconductors} (Princeton University press, Princeton and Oxford, 2013).

\bibitem{Th}
 D.~J.~Thouless, M.~Kohmoto, M.~P.~Nightingale, and M.~den~Nijs, Phys. Rev. Lett. \textbf{49}, 405 (1982).

\bibitem{Von}
K.~v.~Klitzing, G.~Dorda, and M.~Pepper, Phys. Rev. Lett. \textbf{45}, 494 (1980).

\bibitem{hal}
F.~D.~M.~Haldane, Phys. Rev. Lett. \textbf{61}, 2015 (1988).

\bibitem{Kane1}
C.~L.~Kane and E.~J.~Mele, Phys. Rev. Lett. \textbf{95}, 226801 (2005).

\bibitem{Kane2}
C.~L.~Kane and E.~J.~Mele, Phys. Rev. Lett. \textbf{95}, 146802 (2005).

\bibitem{ber} 
A.~B.~Bernevig, S.~C.~Zhang, Phys. Rev. Lett. \textbf{96}, 106802 (2006).


\bibitem{FuKane}
L.~Fu, and C.~L.~Kane, Phys. Rev. B \textbf{76}, 045302 (2007).


\bibitem{moore}
J.~E.~Moore and L.~Balents, Phys. Rev. B \textbf{75}, 121306(R) (2007). 

\bibitem{Altland}
A.~Altland and M.~R.~Zirnbauer, Phys. Rev. B \textbf{55}, 1142 (1997). 

\bibitem{Kruthoff}
J.~Kruthoff, J.~de Boer, J.~van Wezel, C.~L.~Kane and R.~Jan.~Slager, Phys. Rev. X,\textbf{7}, 041069 (2017). 

\bibitem{sch}
A.~P.~Schnyder, S.~Ryu, A.~Furusaki, and A.~W.~W.~Ludwig, Phys. Rev. B \textbf{78}, 195125 (2008).

\bibitem{kit}
A.~Kitaev, AIP Conference Proceedings  \textbf{1134}, 22 (2009), https://aip.scitation.org/doi/pdf/10.1063/1.3149495.

\bibitem{Fu3D}
L.~Fu, C.~L.~Kane and E.~J.~Mele, Phys. Rev. Lett., \textbf{98}, 106803 (2007).

\bibitem{berscience}
B.~A.~Bernevig, T.~L.~Hughes, and S.~-C.~Zhang, Science \textbf{314}, 1757 (2006).

\bibitem{Konig}
M.~K\"onig, S.~Wiedmann, C.~Brüne, A.~Roth, H.~Buhmann, L.~W.~Molenkamp, X.~L.~Qi,  and S.~C.~Zhang, Science, \textbf{318},766 (2007).

\bibitem{Hsieh3D}
D.~Hsieh, D.~Qian, L.~Wray, Y.~Xia, Y.~S.~Hor, R.~J.~Cava, M.~Z.~Hasan, Nature, \textbf{452}, 970, (2008).

\bibitem{Gurarie1}
V.~Gurarie, Phys. Rev. B, \textbf{83}, 085426, (2011).

\bibitem{Gurarie2}
A.~M.~Essin,V.~Gurarie, Phys. Rev. B, \textbf{84}, 125132, (2011).

\bibitem{Wang}
  Z.~Wang and S.~C.~Zhang, Phys. Rev. X, \textbf{2}, 031008 (2012); Z.~Wang, X.-L. Qi, and S.~C.~Zhang, Phys. Rev. B \textbf{85}, 165126 (2012).

\bibitem{rac}
S.~Rachel, Reports on Progress in Physics  \textbf{81}, 116501 (2018).

\bibitem{Hohen}
M.~Hohenadler and F.~F.~Assaad, Journal of Physics:Condensed Matter \textbf{25}, 143201 (2013).

\bibitem{Dag}
M.~Daghofer and M.~Hohenadler, Phys. Rev. B  \textbf{89}, 035103 (2014).

\bibitem{Capponi1}
S.~Capponi and A.~M.~L\"auchli, Phys. Rev. B \textbf{92}, 085146 (2015).

\bibitem{Capponi2}
S.~Capponi, Journal of Physics: Condensed Matter, \textbf{29}, 043002 (2016).

\bibitem{Ruegg}
A.~R\"uegg and G.~A.~Fiete, Phys. Rev. B, \textbf{84}, 201103, (2011).

\bibitem{Amaricci}
A.~Amaricci, J.~C.~Budich, M.~Capone, B.~Trauzettel and G.~Sangiovanni, Phys. Rev. Lett., \textbf{114}, 185701, (2015).

\bibitem{RachLeHur}
S.~Rachel and K.~Le~Hur, Phys. Rev. B  \textbf{82}, 075106 (2010).

\bibitem{Zheng}
D.~Zheng, G.~M~.Zhang and C.~Wu, Phys. Rev. B \textbf{84}, 205121 (2011).

\bibitem{Hohen2}
M.~Hohenadler, Z.~Y.~Meng, T.~C.~Lang, S.~Wessel, A.~Muramatsu and F. F.~Assaad, Phys. Rev. B \textbf{85}, 115132 (2012).

\bibitem{Raghu}
S.~Raghu, X.~L.~Qi, C.~Honerkamp and S.~C.~Zhang, Phys. Rev. Lett. \textbf{100}, 156401 (2008).

\bibitem{Neupert}
T.~Neupert, L.~Santos, C.~Chamon and C.~Mudry, Phys. Rev. Lett. \textbf{106}, 236804 (2011).

\bibitem{Mac}
J.~Maciejko and G.~A.~Fiete, Nature Phys. 11 , 385, (2015).

\bibitem{Gonz}
D.~Gonz\'alez-Cuadra, P.~R.~Grzybowski, A.~Dauphin, and M.~Lewenstein, Phys. Rev. Lett. \textbf{121}, 090402 (2018).

\bibitem{Hohen3}
C.~Chen, X.~Y.~Xu, Zi~Y.~Meng, and M.~Hohenadler, Phys. Rev. Lett. \textbf{122}, 077601 (2019).

\bibitem{Can}
L.~M.~Cangemi, A.~S.~Mishchenko, N.~Nagaosa, V.~Cataudella and G.~ De Filippis,  arXiv:1903.11659.

\bibitem{mona}
M.~M.~M\"oller, G.~A.~Sawatzky, M.~Franz, and M.~Berciu, Nature Communications \textbf{8}, Article number: 2267 (2017). 


\bibitem{prok}
I.~S.~Tupitsyn and N.~V.~Prokof'ev,  arXiv:1809.01258 [cond-mat.str-el]
    
\bibitem{sen}
  D.~S\'en\'echal, D.~Perez, and D.~Plouffe, Phys. Rev. B \textbf{66}, 075129 (2002).
    
\bibitem{mahan}
 G.~D.~Mahan, {\it Many-particle physics}, New York, Plenum Press, 1981.   

\bibitem{fetter}
  A.~L.~Fetter and J.~D.~Walecka, {\it Quantum Theory of Many Particle System}, McGraw-Hill Book Company, New York, 34, 1971.

\bibitem{KiMi}  K.~A.~Kikoin and A.~S.~Mishchenko, ZETF {\bf 104}, 3810 (1993) [JETP {\bf 77}, 828 (1993)].

\bibitem{Citro} G.~De~Filippis, V.~Cataudella,  R.~Citro,  C.~A.~Perroni, A.~S.~Mishchenko, and N.~Nagaosa, Europhys.\ Lett., {\bf 91}, 47007 (2010).

  
  
\end{thebibliography}

%
%

\end{document}